# Special Issue on Spacetime Structure and Electrodynamics

Editors: *Wei-Tou Ni, Sperello di Serego Alighieri, Jonathan Kaufman, Brian Keating*

In the last two decades we have seen important mutual stimulations between the community working on electrodynamics of continuous media and the community working on spacetime structure. This is highlighted by the publication of two important monographs from two communities: *Foundations of Classical Electrodynamics* by F. W. Hehl and Yu. N. Obukhov (Birkhäuser, Boston 2003) and *Differential Forms in Electromagnetics* by I. V. Lindell (IEEE Press-Wiley, Piscataway, NJ 2004; see also a new book "Multiforms, Dyadics, and Electromagnetic Media" by the same author in 2015). Starting around 1960, magnetoelectric effects and magnetoelectric media have been a focus of study. Somewhat later, the constitutive tensor density framework was used to construct spacetime structure theoretically and empirically. Earlier, in putting Maxwell equations into a form compatible with general relativity, Einstein noticed that the Maxwell equations can be formulated in a form independent of the metric gravitational potential in 1916; only the constitutive tensor density depends on the metric. This was followed by further stimulating and clarifying works of Weyl in 1918, Murnaghan in 1921, Kottler in 1922 and Cartan in 1923. Year 2016 is the centennial of the start of this historical development.

International Journal of Modern Physics D of World Scientific is celebrating the occasion by publishing this special issue on Spacetime Structure and Electrodynamics. We have invited researchers to submit original research articles that would contribute to a better understanding of the phenomena inherent to spacetime structure and electrodynamic media.

Maxwell equations are naturally formulated in differential manifold without any additional geometrical structure with constitutive tensor density connecting the electromagnetic excitation density and electromagnetic field intensity. The gadget is that this constitutive tensor could be accounting for the electrodynamics in a continuous medium, and it could also be accounting for the electrodynamics of spacetime (vacuum) in a gravitational field. Different theories of gravity could have different constitutive tensors. General relativity has metric constitutive tensor density. Other theories may have different ones. How do we find the right one? *The guiding principle is Weak Equivalence Principle (WEP) for photons/electromagnetic wave packets.* From the inspiration of Galileo Weak Equivalence Principle (GWEP, Galileo's EP or simply WEP) for test bodies and that photon in vacuum (spacetime) only travels with limiting velocity, the WEP for photons/wave packets can simply be stated as "that the spacetime trajectory of light in a gravitational field depends only on its initial position and direction of propagation, does not depend on its frequency (energy) and polarization."

*This is equivalent to nonbirefringence of light propagation.* From this simple equivalence principle or nonbirefringence, it has been proved in this framework that the constitutive tensor density must be of core metric form with an additional axion (pseudoscalar) and an additional dilaton (scalar) degrees of freedom. The additional axion degree of freedom would give rise to Cosmic Polarization Rotation (CPR) for electromagnetic wave propagation. The empirical verification of WEP for photons/electromagnetic wave packets is very good: from cosmic electromagnetic wave propagation in various direction, it is verified to the order of $10^{-38}$, that is, to $10^{-4} \times$ $O([M_{Higgs}/M_{Planck}]^2$. There are also empirical constraints on the cosmic axions and cosmic dilatons but mild.

In parallel to our call for papers, Osservatorio Astrofisico di Arcetri of INAF organized a workshop on "Cosmic Polarization Rotation: from Galilean Principles to Cosmology." The aim of this workshop is to get together theorists, who develop models leading to such cosmic polarization rotation, and experimentalists, who search for CPR using a variety of complementary methods, in order to discuss current results and future directions. The workshop has taken place at the Villa Il Gioiello on the Arcetri hill near Florence, where Galileo, the first scientists to test the equivalence principle, spent the last 10 years of his life. The orientation of polarization appears to be the most stable property of photons. However, changes in the polarization angle of photons traveling over cosmological distances are foreseen, for example, if fundamental physical principles, such as the Einstein Equivalence Principle, are violated. This Workshop is also in celebration of the Centennial of the theory of General Relativity, developed mostly by Albert Einstein in 1915. In fact CPR has not yet been detected (current upper limits are of the order of 1 degree) and a null CPR results is a strong test of the Einstein Equivalence Principle, on which General Relativity is based.

This special issue contains a total of eighteen articles – nine from call for papers and nine from selected papers presented in the CPR workshop. It started with two reviews on spacetime structure and electrodynamics of continuous media from premetric formulation of electrodynamics by Ni and by Hehl, Itin and Obukhov followed by two reprints – one gives the first serious proposal of Abelian axion and its CPR effect, the other is a festschrift of Brans expounding related work of Brans and Dicke together with their influence and some following developments, and five more articles from call of papers. Hammond in his article searches for a relation between stringy Maxwell charge and the magnetic dipole moment. Denisov, Ilyina and Sokolov search for the nonlinear vacuum electrodynamic influence on the spacetime structure. Kruglov calculates the universe acceleration due to nonlinear electromagnetic field. Stoica expounds Kaluza theory with zero-length extra dimension. Since all these efforts may facilitate ways to explore the origins of gravity, and since axionic medium does exist, efforts to find corresponding media with dilatons and antisymmetric skewons will be warranted. Since Ohm's law is not manifestly relativistic covariance, it would be nice to see a manifestly covariance form. Starke and Schober review the explicit covariance of Ohm's law.

Focusing on CPR, Alexander, and Caldwell have accounted for various recent theories/models which leads to observable CPRs. Gubitosi addresses the issue of disentangling CPR/birefringence from standard physics in CMB measurements and distinguishing among production mechanisms,

Contaldi, and Gruppuso have analyzed Planck CMB polarization data and give good constraints on the uniform CPR rotation angle of about 1° (0.02 rad). Combined with other observations, the constraints on time variation is of the same order of magnitude as the space fluctuations. It may mean our universe is pretty in dynamical equilibrium as far as $\varphi$ is concerned. Therefore the measurement of its spatial fluctuations is even more important.

In the original gravity model with a pseudoscalar, the natural coupling strength $\phi$ is of order 1. However, the isotropy of our observable universe to $10^{-5}$ may lead to a change of $\Delta\varphi$ over cosmological distance scale $10^{-5}$ or smaller. Hence, observations to test and measure $\Delta\varphi$ to $10^{-6}$ are very significant. A positive result may indicate that our patch of inflationary universe has a 'spontaneous polarization' in the fundamental law of electromagnetic propagation influenced by neighboring patches and by a determination of this fundamental physical law we could 'observe' our neighboring patches.

For measurement reaching a polarization angle of $10^{-5}$-$10^{-6}$, calibration angle accuracy is very important. Galluzzi, M. Massardi *et al.* have looked into the properties of polarimetric multi-frequency radio sources related to telescope calibration issues. Kaufman *et al.* looked into the potential of using the Crab Nebula as a high precision calibrator for CMB polarimeters. de Bernardis and Masi discussed possible improvement in the near future on temperature sensitivity and polarization calibration and conclude that: (i) the Large Scale Polarization Explorer (LSPE, http://planck.roma1.infn.it\lspe) on stratospheric balloon platform is expected to constrain the tensor-to-scalar ratio *r* for B-modes with an error $\sigma_r < 0.01$ and the final survey sensitivity for CPR $\leq 0.1°$ ($1\sigma$); (ii) Cosmic Origins Explorer (COrE, a proposal for ESA's M4 space mission) is expected to measure CPR very well, with a final uncertainty in the CPR $\leq 0.01°$ ($1\sigma$) (that is better than $2 \times 10^{-4}$ [rad]), due to the accuracy of calibration and very high sensitivity.

We thank all authors for having collaborated for this special issue, reviewers for elaborate and punctual work, and Kah-Fee Ng, the IJMPD desk editor, for his efficient help in the success of the publication.

# Special IJMPD Issue — Spacetime Structure and Electrodynamics

Editors: Wei-Tou Ni, Sperello di Serego Alighieri,
Jonathan Kaufman and Brian Keating

## Table of Contents with publication information and arXiv numbers

*Phys. D* **25** (2016) 1640010; arXiv:1409.3723

Inflationary birefringence and baryogenesis, Stephon Alexander, *Int. J. Mod. Phys. D* **25** (2016) 1640013; arXiv:1604.00703

Cosmic parity violation due to a flavor-space locked gauge field, Robert Caldwell, 1640011

Disentangling birefringence from standard physics in CMB measurements and distinguishing among production mechanisms, Giulia Gubitosi, *Int. J. Mod. Phys. D* **25** (2016) 1640006

Cosmological birefringence constraints from the Planck 2015 CMB likelihood, Alessandro Gruppuso, *Int. J. Mod. Phys. D* **25** (2016) 1640007

Imaging cosmic polarization rotation, Carlo R. Contaldi, *Int. J. Mod. Phys. D* **25** (2016) 1640014; arXiv:1510.02629

The polarimetric multi-frequency radio sources properties, Vincenzo Galluzzi and Marcella Massardi, *Int. J. Mod. Phys. D* **25** (2016) 1640005;

Polarization of extragalactic radio sources: CMB foregrounds and telescope calibration issues, Marcella Massardi, Vincenzo Galluzzi, Rosita Paladino, and Carlo Burigana, *Int. J. Mod. Phys. D* **25** (2016) 1640009

Using the Crab Nebula as a high precision calibrator for cosmic microwave background polarimeters, Jonathan Kaufman, David Leon and Brian Keating, *Int. J. Mod. Phys. D* **25** (2016) 1640008; arXiv:1602.01153

Cosmic microwave background and cosmic polarization rotation: An experimentalist view, Paolo de Bernardis, Silvia Masi, *Int. J. Mod. Phys. D* **25** (2016) 1640012